\documentclass[11pt,letterpaper]{article}
\usepackage[margin=1in]{geometry}
\usepackage[T1]{fontenc}
\usepackage{amsmath,amssymb}
\usepackage{booktabs}
\usepackage{natbib}
\usepackage[hidelinks,hyperfootnotes=false]{hyperref}
\setcitestyle{authoryear,square,citesep={,},aysep={,},yysep={,}}
\date{}
\title{One Release for All: Private Simple Linear Regression
Via Improved Simplex Mapping}
\author{%
Sasmita Harini S\thanks{Work done while an undergraduate student at the Indian Institute of Science.}\\
Microsoft India\\
\texttt{t-sasmitas@microsoft.com}
\and
Anshoo Tandon\\
CDPG, Indian Institute of Science\\
\texttt{anshoo.tandon@gmail.com}
}
\hypersetup{pdftitle={One Release for All: Private Simple Linear Regression Via Improved Simplex Mapping},pdfauthor={Sasmita Harini S and Anshoo Tandon}}

\begin{document}
\maketitle
\begin{abstract}
We present a single-release extension of DP-RSS \citep{harini2026refined}
for private simple linear regression when both input and output are bounded in
$[0,1]$. Rather than privately releasing several related sufficient-statistic
estimates, we encode the required information in one unit-sensitivity vector
and apply the Laplace mechanism once at the full privacy budget. The required
least-squares sufficient statistics are recovered by deterministic
post-processing and used directly in an ordinary least-squares solve. We
obtain variance reductions of $1.67$--$2.67\times$ relative to DP-RSS for several
statistics. Experiments show that the benefit is most pronounced in the
high-privacy regime, while the methods converge as privacy noise becomes
negligible.
\end{abstract}

\section{Introduction}
Differential privacy provides a formal framework for releasing statistics and
learning models from sensitive data
\citep{dwork2006calibrating,dwork2014algorithmic}. Regression is a useful
setting for studying private statistical release because ordinary least squares
depends on a small collection of algebraic sufficient statistics. Existing
private regression methods include private empirical risk minimization
\citep{kifer2012private,bassily2014private}, functional perturbation
\citep{zhang2012functional}, perturbation of ordinary-least-squares statistics
\citep{sheffet2017differentially}, adaptive sufficient-statistic methods such as
\textsc{AdaSSP} \citep{wang2018PrivateLinearRegression}, and methods specifically
designed for simple linear regression
\citep{alabi2022differentially,amin2023easyDP}. More recent work studies
data-dependent sufficient-statistic perturbation \citep{ferrando2025private}
and the statistical cost of privacy more generally
\citep{cai2021cost,kamath2023trilemma}. Another prominent approach is DP-SGD,
which clips per-sample gradients and injects calibrated noise during iterative
optimization \citep{bassily2014private,abadi2016deep,cai2021cost}. These works
provide foundational privacy--utility guarantees for private optimization,
while in practice DP-SGD typically requires careful choices of clipping
thresholds and other optimization hyperparameters.

Independently, and concurrently with our work, Takakura et
al. developed the
B\'{e}zier mechanism, which leverages Bernstein/B\'{e}zier polynomial structure
for private variance and covariance estimation in the add-remove
model~\citep{takakura2025optimalvariancecovarianceestimation}.
Takakura et al. also mention potential extension of their construction to
multivariate tensor products. In particular, $\{1-y,y\}$ is a standard
degree-one Bernstein response basis, not a novel decomposition. We do not
claim priority over the shared Bernstein construction or optimal variance
factors among private mechanisms. Here we study the recovery of the five OLS
sufficient statistics for simple linear regression and their use in an
unregularized OLS solve.

Our starting point is DP-RSS \citep{harini2026refined}, which uses algebraic
structure to improve private simple linear regression, building on recent work
on private mean estimation \citep{kulesza2024,fitzsimons2024}. We ask "can the
sufficient-statistic workload of simple linear regression be packaged into a
single unit-sensitivity private vector release?".

We answer this question for simple linear regression with one feature and one
response, where both variables lie in $[0,1]$. For each record, we construct
six nonnegative coordinates that sum to one. The aggregate six-vector
therefore has unit $\ell_1$ sensitivity under add-remove adjacency. Here,
add-remove adjacency means that two datasets differ by one record, either
through the addition or removal of that record. We then apply one
vector-valued Laplace mechanism at the full privacy budget and recover all
sufficient statistics required by ordinary least squares through deterministic
post-processing.

This construction illustrates a broader design principle i.e when an estimator
depends on several related statistics, a low-sensitivity joint representation
can avoid the privacy-budget splitting required when the statistics are
released separately.

\begin{figure}[t]
\small
\refstepcounter{figure}\label{alg:single}
\setlength{\parskip}{0pt}
\hrule\smallskip
\noindent\textbf{Algorithm \thefigure}\quad Single-release algorithm for private simple linear regression.\par
\smallskip\hrule\smallskip
\textbf{Require:} $D=\{(x_i,y_i)\}_{i=1}^{n}$, $(x_i,y_i)\in[0,1]^2$; $\epsilon$.
\begin{enumerate}
\setlength{\itemsep}{1pt}\setlength{\parsep}{0pt}
\item Initialize $B\gets[0,0,0,0,0,0]$.
\item \textbf{for} each $(x_i,y_i)\in D$ \textbf{do}
\item\hspace*{1em}%
  $\begin{aligned}[t]
  \phi_i\gets\bigl(& (1-x_i)^2(1-y_i),\;2x_i(1-x_i)(1-y_i),\;x_i^2(1-y_i),\\
  & (1-x_i)^2y_i,\;2x_i(1-x_i)y_i,\;x_i^2y_i\bigr)
  \end{aligned}$
\item\hspace*{1em}$B\gets B+\phi_i$.
\item \textbf{end for}
\item Draw $Z_1,\ldots,Z_6\overset{\mathrm{iid}}\sim\mathrm{Lap}(1/\epsilon)$ and set
$\widetilde B[j]\gets B[j]+Z_{j+1}$ for $j=0,\ldots,5$.
\item Recover $\hat n\gets\sum_{j=0}^{5}\widetilde B[j]$,
$\hat S_{x^2}\gets\widetilde B[2]+\widetilde B[5]$,
$\hat S_{xy}\gets\tfrac12\widetilde B[4]+\widetilde B[5]$.
\item Recover $\hat S_x\gets\tfrac12(\widetilde B[1]+\widetilde B[4])+\widetilde B[2]+\widetilde B[5]$,
$\hat S_y\gets\widetilde B[3]+\widetilde B[4]+\widetilde B[5]$.
\item \textbf{Return:} $\displaystyle
(\hat\alpha,\hat\beta)^\top\gets
\begin{pmatrix}\hat n&\hat S_x\\\hat S_x&\hat S_{x^2}\end{pmatrix}^{-1}
\begin{pmatrix}\hat S_y\\\hat S_{xy}\end{pmatrix}$
\end{enumerate}
\medskip\hrule
\end{figure}

\section{Methodology}
Consider a dataset $D=\{(x_i,y_i)\}_{i=1}^{n}$ of $n$ rows, where
$x_i\in\mathbb R$ is a numerical feature and $y_i\in\mathbb R$ is the
corresponding response. We consider the simple linear regression model
$y_i=\alpha+\beta x_i+\eta_i$, where $\eta_i\sim\mathcal N(0,\sigma^2)$
represents observation noise. Following the common bounded-data assumption
in differentially private regression \citep{harini2026refined}, we assume that
the ranges of both variables are publicly known and, without loss of
generality, normalize them such that $x_i,y_i\in[0,1]$ for all $i$. This
normalization is performed deterministically before the private computation
and therefore does not affect the differential privacy guarantee. The resulting
statistics and regression coefficients can be mapped back to the original
scale after the private computation.

Private linear regression based on ordinary least squares is typically
realized through private release of the sufficient statistics $n$,
$S_x=\sum_i x_i$, $S_y=\sum_i y_i$, $S_{x^2}=\sum_i x_i^2$, and
$S_{xy}=\sum_i x_i y_i$. Given these statistics, the exact
ordinary-least-squares (OLS) estimators of the slope and intercept are
\[
\hat\beta_{\mathrm{OLS}}=\frac{nS_{xy}-S_xS_y}{nS_{x^2}-S_x^2},
\qquad
\hat\alpha_{\mathrm{OLS}}=\frac{S_y-\hat\beta_{\mathrm{OLS}}S_x}{n}.
\]
Thus, our private mechanism only needs to recover the five quantities $n$,
$S_x$, $S_y$, $S_{x^2}$, and $S_{xy}$, after which the OLS coefficients are
obtained deterministically by post-processing.

For each record $(x,y)$, we define a six-dimensional vector as follows:
\[
\begin{aligned}
\phi(x,y)=\bigl(& (1-x)^2(1-y),\;2x(1-x)(1-y),\;x^2(1-y),\\
& (1-x)^2y,\;2x(1-x)y,\;x^2y\bigr).
\end{aligned}
\]
The six coordinates are nonnegative and sum to one. Hence, for
$B(D)=\sum_i\phi(x_i,y_i)$, we have $\Delta_1(B)=1$ under add-remove
adjacency. We therefore release
\[
\widetilde B=B(D)+Z,\qquad Z_j\overset{\mathrm{iid}}\sim\mathrm{Lap}(1/\varepsilon).
\]
This single vector-valued Laplace mechanism provides pure
$\varepsilon$-differential privacy, i.e., $(\varepsilon,0)$-DP.
Algorithm~\ref{alg:single} gives the complete algorithm. It requires one pass
over the data to accumulate the six-dimensional statistic, followed by a
fixed-size Laplace release and deterministic recovery of the ordinary least-squares
statistics. The total computational complexity is therefore $O(n)$ with
$O(1)$ additional memory.

\subsection{Construction Details}
\paragraph{Basis functions and unit-sum property.}
The six-dimensional representation uses quadratic basis functions
$q_0(x)=(1-x)^2$, $q_1(x)=2x(1-x)$, $q_2(x)=x^2$ that form a partition of
unity: $q_0(x)+q_1(x)+q_2(x)=1$. These encode
$x=\tfrac12q_1(x)+q_2(x)$ and $x^2=q_2(x)$. By multiplying each basis
function by $(1-y)$ and $y$, we obtain six coordinates
in the order used in Algorithm~\ref{alg:single}:
$\phi_k(x,y)=q_k(x)(1-y)$ and $\phi_{k+3}(x,y)=q_k(x)y$ for $k=0,1,2$,
ensuring $\sum_{j=0}^{5}\phi_j(x,y)=1$ and unit $\ell_1$ sensitivity under
add-remove adjacency.

\paragraph{Recovery identities and weight vectors.}
From $\widetilde B=B(D)+Z$ where
$Z_j\overset{\mathrm{iid}}\sim\mathrm{Lap}(1/\varepsilon)$, we recover all
sufficient statistics via linear combinations of its components. Each recovery
formula defines a recovery weight vector $w$: a 6D vector whose entries are
the coefficients. For example, the recovery of $S_x$ is
$\hat S_x=\tfrac12\widetilde B[1]+\widetilde B[2]+\tfrac12\widetilde B[4]+\widetilde B[5]$,
giving weight vector $w_{S_x}=(0,\tfrac12,1,0,\tfrac12,1)$. The complete recovery
identities are $\hat n=\sum_{j=0}^{5}\widetilde B[j]$,
$\hat S_{x^2}=\widetilde B[2]+\widetilde B[5]$,
$\hat S_{xy}=\tfrac12\widetilde B[4]+\widetilde B[5]$, and
$\hat S_y=\widetilde B[3]+\widetilde B[4]+\widetilde B[5]$.

\section{Variance Analysis and Reduction}
Since $Z_j\sim\mathrm{Lap}(1/\varepsilon)$ has variance
$\operatorname{Var}[Z_j]=2/\varepsilon^2$ and components are independent, the
variance of a weighted combination of noisy statistics is
\[
\operatorname{Var}[\hat S]
=\operatorname{Var}\!\left[\sum_{j=0}^{5}w_j\widetilde B[j]\right]
=\sum_{j=0}^{5}w_j^2\frac{2}{\varepsilon^2}
=\frac{2}{\varepsilon^2}\|w\|_2^2
\]
and $\|w_{S_x}\|_2^2=\tfrac52$, $\|w_n\|_2^2=6$,
$\|w_{S_{x^2}}\|_2^2=2$, $\|w_{S_y}\|_2^2=3$,
$\|w_{S_{xy}}\|_2^2=\tfrac54$.

These weight norms produce the mechanism-level variances shown in
Table~\ref{tab:variance}, demonstrating substantial variance reduction over
DP-RSS. The count statistic is unchanged because both methods use six unit
weights. All other statistics achieve lower variance because their recovery
weight vectors have smaller squared norms. This is the fundamental advantage
of our approach: by encoding multiple related sufficient statistics into a
single low-sensitivity representation, we avoid splitting the privacy budget,
making the weights more efficient than methods that release statistics
separately.

\begin{table}[t]
\centering
\caption{Exact variance of recovered sufficient statistics.}
\label{tab:variance}
\begin{tabular}{lccc}
\toprule
Statistic & DP-RSS & Our method & Reduction\\
\midrule
$\hat n$ & $12/\varepsilon^2$ & $12/\varepsilon^2$ & $1.00\times$\\
$\hat S_{x^2}$ & $20/(3\varepsilon^2)$ & $4/\varepsilon^2$ & $1.67\times$\\
$\hat S_{xy}$ & $20/(3\varepsilon^2)$ & $5/(2\varepsilon^2)$ & $2.67\times$\\
$\hat S_x$ & $32/(3\varepsilon^2)$ & $5/\varepsilon^2$ & $2.13\times$\\
$\hat S_y$ & $32/(3\varepsilon^2)$ & $6/\varepsilon^2$ & $1.78\times$\\
\bottomrule
\end{tabular}
\end{table}

\section{Experiments and Discussion}
We evaluate our method on two synthetic data settings and compare our method
with DP-RSS \citep{harini2026refined}, \textsc{DiffPrivLib}
\citep{holohan2019diffprivlib}, \textsc{BinAgg} \citep{lin2025dplrsynthetic},
and \textsc{AdaSSP} \citep{wang2018PrivateLinearRegression}. We did not include
DP-TheilSen \citep{alabi2022differentially} in the comparisons because it was
shown to consistently under-perform DP-RSS in \citet{harini2026refined}. For
DiffPrivLib, we use the corrected implementation described by
\citet{cebere2026bug} to ensure a fair comparison. Results are averaged over
200 Monte Carlo seeds using an 80/20 train/test split.

\paragraph{Privacy scope of preprocessing.}
The privacy guarantee assumes a fixed public choice of features and fixed
public bounds for normalization. Any training-data-dependent feature selection
or empirical sample normalization is not covered by the mechanism's reported
privacy budget $\varepsilon$ and requires separate privacy accounting.
Deterministic preprocessing preserves the stated guarantee here only when its
parameters are fixed independently of the private training data.

\paragraph{Dataset Type 1: Linear model and sample-size effects.}
We generate
\[
x_i\sim\mathrm{Uniform}(0,1),\qquad
y_i=\mathrm{clip}\bigl(0.2+0.6x_i+\mathcal N(0,0.03^2),0,1\bigr).
\]
Table~\ref{tab:linear} shows that our method consistently achieves the lowest
test error across all sample sizes. The largest improvement occurs in the
smallest-sample regime, where the proposed method reduces error by
approximately $38\%$ relative to DP-RSS and provides still larger gains over
DiffPrivLib, BinAgg, and \textsc{AdaSSP}. As the sample size grows, all
sufficient-statistic-based approaches become increasingly similar, and the
advantage over DP-RSS becomes negligible.

\begin{table}[t]
\centering
\caption{Dataset Type 1. Test MSE ($\times10^{-3}$) at $\varepsilon=1$.}
\label{tab:linear}
\begin{tabular}{rrrrrr}
\toprule
$n$ & Our method & DP-RSS & DiffPrivLib & BinAgg & AdaSSP\\
\midrule
500 & 0.975 & 1.583 & 2.426 & 6.868 & 16.445\\
2,000 & 0.901 & 0.938 & 1.004 & 2.005 & 4.687\\
10,000 & 0.900 & 0.901 & 0.904 & 1.115 & 1.026\\
\bottomrule
\end{tabular}
\end{table}

\paragraph{Dataset Type 2: Nonlinear relationships and privacy effects.}
To study robustness, we consider
\[
y_i=\mathrm{clip}\bigl(x_i^2+\mathcal N(0,0.03^2),0,1\bigr)
\]
for the quadratic setting and
\[
y_i=\mathrm{clip}\bigl(0.5+0.4\sin(2\pi x_i)+\mathcal N(0,0.03^2),0,1\bigr)
\]
for the sinusoidal setting. Both methods continue to fit the same linear
ordinary least-squares model, so these experiments isolate the effect of the
private release mechanism when the data-generating relationship itself is
nonlinear.

The results in Table~\ref{tab:nonlinear} show a consistent privacy-dependent
pattern across both nonlinear relationships. Under the strongest privacy
constraint, $\varepsilon=0.1$, our method improves test MSE relative to DP-RSS
by approximately $10\%$ for the quadratic relationship and $3\%$ for the
sinusoidal relationship. The gains over DiffPrivLib are approximately $30\%$
and $5\%$, respectively, while substantially larger improvements are observed
relative to BinAgg and AdaSSP. As the privacy budget increases, the difference
between our method and DP-RSS rapidly contracts to below $1\%$, with the
methods becoming nearly indistinguishable at $\varepsilon=1.0$. Thus, the
advantage of the single-release construction persists under nonlinear model
misspecification, but primarily affects the privacy-induced component of error
rather than the approximation error caused by fitting a linear model to
nonlinear data.

\begin{table}[t]
\centering
\caption{Dataset Type 2: Nonlinear models. Test MSE ($\times10^{-3}$).}
\label{tab:nonlinear}
\small
\begin{tabular}{lrrr@{\hspace{18pt}}rrr}
\toprule
&\multicolumn{3}{c}{Quadratic}&\multicolumn{3}{c}{Sinusoid}\\
\cmidrule(lr){2-4}\cmidrule(lr){5-7}
Method & $\varepsilon{=}0.1$ & $0.5$ & $1.0$ & $\varepsilon{=}0.1$ & $0.5$ & $1.0$\\
\midrule
Our method & 6.757 & 6.628 & 6.623 & 32.374 & 32.266 & 32.262\\
DP-RSS & 7.490 & 6.652 & 6.628 & 33.476 & 32.304 & 32.270\\
DiffPrivLib & 9.646 & 6.717 & 6.654 & 33.955 & 32.340 & 32.271\\
BinAgg & 18.888 & 8.649 & 7.223 & 47.241 & 36.137 & 33.761\\
AdaSSP & 46.755 & 13.369 & 8.138 & 82.884 & 41.607 & 33.749\\
\bottomrule
\end{tabular}
\end{table}

\paragraph{Future Directions.}
The present construction is deliberately specialized to simple linear
regression with one bounded feature and one bounded response. Its key
requirement is a compact nonnegative representation whose coordinates sum to
one while retaining the information needed to recover the relevant sufficient
statistics. Extending this principle to higher-dimensional regression without
introducing prohibitive vector dimension or high sensitivity remains an open
direction.

\section{Conclusion}
We presented a single-release extension of DP-RSS for private simple linear
regression. By encoding the information required by ordinary least squares
into a unit-sensitivity six-dimensional aggregate, our method releases the
relevant sufficient statistics using the full privacy budget. Exact
calculations give $1.67$--$2.67\times$ lower variance for the non-count
sufficient statistics relative to DP-RSS.

The empirical results show that this mechanism-level improvement is most
useful in high-privacy regimes. Our method achieves the lowest reported test
MSE across all the experiments. As privacy noise decreases via larger $n$ or
larger $\varepsilon$, the methods become increasingly similar and the
remaining error is increasingly dominated by sampling and approximation
effects.

Overall, the results support that carefully packaging related sufficient
statistics into a single low-sensitivity vector release can reduce
privacy-induced noise without changing the underlying ordinary least-squares
estimator.

\end{document}